\documentclass[amsmath,amssymb,a4paper]{revtex4}
\usepackage{bm}
\usepackage{amsfonts}

\newtheorem{theorem}{Theorem}

\newtheorem{lemma}{Lemma}

\newenvironment{proof}
{\par\noindent {\bf Proof}}
{\par\strut\hfill$\square$\par\vskip 0.5cm}

\newcommand{\be}{\begin{eqnarray}}
\newcommand{\ee}{\end{eqnarray}}

\begin{document}

\title{Transition records of stationary Markov chains}
\author{Jan Naudts}
\author{Erik Van der Straeten}
\affiliation{Departement Fysica, Universiteit Antwerpen, Groenenborgerlaan 171, 2020 Antwerpen, Belgium}
\email[E-mail]{jan.naudts@ua.ac.be, erik.vanderstraeten@ua.ac.be}

\begin{abstract}
In any Markov chain with finite state space the distribution of transition records
always belongs to the exponential family. This observation is used to prove
a fluctuation theorem, and to show that the dynamical entropy
of a stationary Markov chain is linear in the number of steps.
Three applications are discussed.
A known result about entropy production is reproduced.
A thermodynamic relation is derived for equilibrium systems with Metropolis dynamics.
Finally, a link is made with recent results concerning a one-dimensional polymer model.
\end{abstract}


\maketitle

\paragraph* {Introduction}

The fluctuation theorem, discovered in 1993 by Evans, Cohen and Morriss \cite {ECM93}, was the start of a series of
new developments in the study of stationary non-equilibrium systems (see for example \cite {ESR05} for an overview).
A formulation of the fluctuation theorem in stochastic context was given in \cite {LS99,MC99}.
These authors showed that the theorem holds in full generality for any stationary probability
distribution of any Markov process. The proof given here is restricted
to Markov chains with finite state space. In this way technicalities can be avoided.

The method of proof of the present paper is based on the observation that a certain probability distribution,
hereafter called {\sl distribution of transition records,}
always belongs to the exponential family.
From this result the fluctuation theorem follows immediately.
Indeed, Lebowitz and Spohn \cite {LS99} and Maes \cite {MC99} already observed that
a Gibbs distributed random variable is all what is needed to derive a
fluctuation theorem of the Gallavotti-Cohen type \cite {GC95} in a stochastic context.
This point has been further elaborated in \cite {JQZ03}.

The motivation for studying the distribution of transition records was twofold.
At one hand the present authors had noticed \cite {VN05,VN06} that, in the Markov chain
of increments of some model of random walk describing a polymer, this distribution
belongs automatically to the exponential family. At the other hand, it is
a straightforward generalization of a distribution introduced in recent work
of Carati \cite {CA05,CA06}. This author assumed a Poisson distribution
based on the assumption that subsequent visits to macrocells in phase space of
a system of classical mechanics are mutually independent.

\paragraph* {Definitions}

Throughout the paper we consider a fixed state space $\Gamma$, containing a finite number $N$
of states. A Markov chain with state space $\Gamma$ is determined by initial probabilities
$p(x)$, with $x$ in $\Gamma$, and by transition probabilities $w(x,y)$, with $x$ and $y$ in $\Gamma$.
The probabilities $p(x)$ are used as initial values for the equation of motion
\be
p_{t+1}(x)=\sum_{y\in\Gamma}p_t(y)w(y,x).
\ee

The probability distribution $p(x)$ is {\sl stationary} if
\be
p(x)=\sum_{y\in \Gamma}p(y)w(y,x)
\label {beq}
\ee
for all $x\in \Gamma$. A stronger condition is {\sl detailed balance}
\be
p(x)w(x,y)=p(y)w(y,x)
\label {dbc}
\ee
for any pair of states $x,y\in \Gamma$.
Summing this equality over $y$ yields (\ref {beq}).

The $w(x,y)$ are parameters of the Markovian model. However, they are not independent
because of the normalization condition
$\sum_yw(x,y)=1$.
Introduce therefore independent parameters $\theta_{x,y}$ and dependent $\eta_x$ such that
\be
w(x,y)&=&w_0(x,y)e^{-\theta_{x,y}}\qquad\mbox {if }x\not=y\cr
w(x,x)&=&e^{-\eta_x},
\ee
with enabling factors $w_0(x,y)$
satisfying $w_0(x,y)=w_0(y,x)$, and $w_0(x,y)=1$ or $w_0(x,y)=0$.
The parameters $\eta_x$ are considered to be functions of the $\theta_{x,y}$.
For convenience, let $\theta_{x,x}=0$ and $w_0(x,x)=1$
for all $x$.

\paragraph* {A Gibbs distribution for transition records}

The record of transitions $k$ is a sequence of numbers $k_{x,y}$,
one for each pair of states $x,y$, counting how many times
the transition from $x$ to $y$ is contained in a given path of the Markov chain.
Our key quantity is the probability $d_x(k)$
that a path starting in $x$ results in a given record of transitions $k$.

It is not difficult to see that one can write
\be
d_x(k)
&=&c_n(x,k)\prod_{y,z}w(y,z)^{k_{y,z}}\cr
&=&c_n(x,k)\exp\left(-\Psi_\theta(k)\right).
\label {gibbs}
\ee
In this expression, $c_n(x,k)$ counts the number of paths that have the same record of transitions $k$
and that are allowed by the enabling factors $w_0(x,y)$.
The dynamical entropy variable $\Psi_\theta(k)$ is given by
\be
\Psi_\theta(k)=\sum_y\eta_yk_{y,y}+\sum_{y,z}\theta_{y,z}k_{y,z}.
\ee
In particular, (\ref {gibbs}) implies that the distribution $d_x(k)$ belongs to
the curved exponential family.

The special form of (\ref {gibbs}) makes it easy to calculate the following identities
\be
0&=&\frac {\partial\,}{\partial\theta_{u,v}}\sum_kd_x(k)\cr
&=&-\langle k_{u,v}\rangle_x+\frac {w(u,v)}{w(u,u)}\langle k_{u,u}\rangle_x,
\ee
with
\be
\langle k_{u,v}\rangle_x&=&
\sum_k d_x(k)k_{u,v}.
\ee
This can be written as
\be
w(u,u)\langle k_{u,v}\rangle_x=\langle k_{u,u}\rangle_xw(u,v).
\label {wdb}
\ee
The latter result is clearly also valid for $u=v$.
This means that $\langle k_{u,v}\rangle_x$, the average number of transitions from $u$ to $v$,
is proportional to $w(u,v)$, the probability to go from $u$ to $v$.

\paragraph* {Linear production of dynamical entropy}

A path $\gamma=(x_0,x_1,\cdots x_n)$ of the Markov chain has probability
$p(x_0)w(\gamma)$ with
\be
w(\gamma)=w(x_0,x_1)\cdots w(x_{n-1},x_n).
\ee
Let $\gamma_i=x_0$ denote the initial and $\gamma_f=x_n$ the final state of $\gamma$.
The {\sl dynamical entropy} $S^{(n)}_\theta$ of the Markov chain is defined by
\be
S^{(n)}_\theta=-\sum_\gamma p(\gamma_i) w(\gamma)\ln w(\gamma).
\ee
The time-reversed dynamical entropy \cite {GP04} is defined by
\be
\overline S^{(n)}_\theta&=&-\sum_\gamma p(\gamma_i) w(\gamma)\ln w(\overline\gamma),\cr
&=&-\sum_\gamma p(\gamma_f) w(\overline\gamma)\ln w(\gamma),
\ee
where $\overline\gamma$ is the reversed path $\overline\gamma=(x_n,x_{n-1},\cdots x_0)$.

Observe that the dynamical entropy $S^{(n)}_\theta$ can be expressed in terms of the
distributions $d_x(k)$. Indeed one has
\be
S^{(n)}_\theta
&=&-\sum_xp(x)\sum_{y,z}\langle k_{y,z}\rangle_x\ln w(y,z)\cr
&=&\sum_xp(x)\langle\Psi_\theta\rangle_x.
\ee
If $p(x)$ is stationary then this expression can be further simplified. One
expects intuitively that each occurrence of some state $y$ contributes to the
dynamical entropy production with an amount $I_y$, defined by
\be
I_y=-\sum_uw(y,u)\ln w(y,u).
\ee
Indeed, this is the result of Theorem \ref {thm1} below.

The corresponding relations for the time-reversed paths are
\be
\overline S^{(n)}_\theta=\sum_xp(x)\langle\overline\Psi_\theta\rangle_x.
\ee
with $\overline\Psi_\theta(k)=\Psi_\theta(\overline k)$, $\overline k_{x,y}=k_{y,x}$, and
\be
\overline I_y=-\sum_uw(y,u)\ln w(u,y).
\ee

Given a path with record of transitions $k$, the number of occurrences of the state $y$
(neglecting the final state) is $\sum_zk_{y,z}$. The probability distribution of
the latter quantity is the distribution studied by Carati \cite {CA05,CA06}.
Using (\ref {wdb}), the average number of occurrences of state $y$ is
\be
\sum_z\langle k_{y,z}\rangle_x=\frac 1{w(y,y)}\langle k_{y,y}\rangle_x.
\ee
The average entropy produced when leaving $y$ is $I_y$. Hence
\be
\sum_y\frac 1{w(y,y)}\langle k_{y,y}\rangle_x I_y
\ee
is the average production of dynamical entropy per step.
One therefore expects the following result.

\begin {lemma}
Assume $w(x,x)>0$ for all $x$. Then one has
\be
\langle\Psi_\theta\rangle_x&=&\sum_y\frac 1{w(y,y)}\langle k_{y,y}\rangle_x I_y,\\
\langle\overline\Psi_\theta\rangle_x&=&\sum_y\frac 1{w(y,y)}\langle k_{y,y}\rangle_x \overline I_y.
\ee
\end {lemma}

Next, one proves by full induction that

\begin {lemma}
Let $p$ be a stationary probability distribution.
Then
\be
\sum_xp(x)\langle k_{y,y}\rangle_x=nw(y,y)p(y).
\ee
\end {lemma}

Combining the different pieces gives the result

\begin {theorem}
\label {thm1}
Assume $p$ is stationary.
Then
\be
S^{(n)}_\theta&=&n\sum_xp(x)I_x,\\
\overline S^{(n)}_\theta&=&n\sum_xp(x)\overline I_x.
\ee
\end {theorem}

Note that the condition of Lemma 1 that $w(x,x)>0$ for all $x$ is not essential for the theorem
to hold. It can be removed by a limiting procedure.

\paragraph* {Fluctuation theorem}

Without any assumption about the Markov chain, the probability
distribution $d_x(k)$ belongs automatically to the exponential family.
As noted in the introduction, this observation
suffices to derive a fluctuation theorem.

Let $C(x,k)$ denote the class of all paths that start in state $x$ and have
the same record of transitions $k$. The probability of this class is $d_x(k)$.
It is easy to see that all paths of this class have the same final state, denoted
$f(x,k)$. This implies that there is a one-to one-correspondence between paths of $C(x,k)$
and reversed paths belonging to $C(f(x,k),\overline k)$.
Introduce the entropy production variable
\be
W(\gamma)=\ln\frac {p(\gamma_i)w(\gamma)}{p(\gamma_f)w(\overline\gamma)}
\ee
(called {\sl action functional} in \cite {LS99}).
Note that, because of (\ref {gibbs}), $W(\gamma)$ is constant on the class $C(x,k)$,
with value
\be
\ln\frac {p(\gamma_i)}{p(\gamma_f)}
-\Psi_\theta(k)+\Psi_{\theta}(\overline k).
\ee
It is now straightforward to derive the fluctuation theorem

\begin {theorem}
\label {ft}
In any Markov chain with finite state space the entropy production variable satisfies
\be
\frac {\mbox{\rm Prob}(W=K)}
{\mbox{\rm Prob}(W=-K)}=e^K.
\ee
\end {theorem}

\begin {proof}

One has

\begin {widetext}
\be
\mbox{\rm Prob}(W=K)
&\equiv&\sum_\gamma p(\gamma_i)w(\gamma)\delta_{\{W(\gamma),K\}}\crcr
&=&e^K\sum_\gamma p(\gamma_f)w(\overline\gamma)\delta_{\{W(\gamma),K\}}\crcr
&=&e^K\sum_\gamma p(\gamma_i)w(\gamma)\delta_{\{W(\overline\gamma),K\}}\crcr
&=&e^K\sum_\gamma p(\gamma_i)w(\gamma)\delta_{\{W(\gamma),-K\}}\crcr
&=&e^K\mbox{\rm Prob}(W=-K).
\ee
\end {widetext}

\end {proof}

In a more general setting \cite {LS99,MC99}, the same result has only been proved
to hold asymptotically for large times.

\paragraph* {Entropy production}

The standard and the time-reversed dynamical entropies are related by
\be
\overline S^{(n)}_\theta-S^{(n)}_\theta
&=&\langle\overline\Psi_\theta\rangle-\langle\Psi_\theta\rangle\crcr
&=&\langle W\rangle-\sum_\gamma p(\gamma_i)w(\gamma)\ln\frac {p(\gamma_i)}{p(\gamma_f)}.
\ee
Application of the fluctuation theorem then gives
\be
\langle W\rangle
&=&\frac 12\sum_\gamma \left(p(\gamma_i)w(\gamma)-p(\gamma_f)w(\overline\gamma)\right)
\ln \frac {p(\gamma_i)w(\gamma)}{p(\gamma_f)w(\overline\gamma)}\crcr
&=&
\overline S^{(n)}_\theta -S^{(n)}_\theta
+\sum_\gamma p(\gamma_i)w(\gamma)\ln\frac {p(\gamma_i)}{p(\gamma_f)}\crcr
&=&\sum_KK\mbox{\rm Prob}(W=K)\cr
&=&\frac 12\sum_K\mbox{\rm Prob}(W=K)\, K(1-e^{-K})\cr
&\ge&0.
\ee
Of course, this positivity can be directly deduced from the first line of the
above expression.

On the other hand, using Theorem \ref {thm1} and assuming stationarity, it is possible to write
\be
\overline S^{(n)}_\theta -S^{(n)}_\theta
&=&n\sum_xp(x)(\overline I_x-I_x)
=n\Delta S,
\label {timerev}
\ee
with
\be
\Delta S&=&\frac 12\sum_{x,y}\left(p(x)w(x,y)-p(y)w(y,x)\right)\cr
& &\times\ln\frac {p(x)w(x,y)}{p(y)w(y,x)}.
\ee
$\Delta S$ is the entropy production \cite {LS99,GP04,LAW05}.
It vanishes for distributions satisfying the detailed balance condition.

\paragraph* {A one-parameter model of equilibrium states}

We now restrict ourselves to equilibrium, which means that probability distributions
satisfy the detailed balance condition. We start from a symmetric matrix $a_{x,y}$
labelled by $x,y$ in the state space $\Gamma$.
It will be used below to fix the transition probabilities $w(x,y)$ of the model.
We fix a parameter $\xi$, which is the analogue of the
inverse temperature $\beta$. In fact, in the case of Metropolis dynamics the matrix
$a_{x,y}$ is given by
\be
a_{x,y}&=&\max\{H(x),H(y)\},\qquad x\not=y\cr
a_{x,x}&=&-\frac 1\beta\ln\left(ce^{-\beta H(x)}-\sum_y^{\not=x}w_0(x,y)e^{-\beta a_{x,y}}\right),
\ee
where $H(x)$ is the Hamiltonian, $\xi$ equals $\beta$, and $c=\sum_y^{\not=x}w_0(x,y)$ independent of $x$.

Introduce a partition function $\Xi(\xi)$,
a probability distribution $p(x)$, and transition probabilities $w(x,y)$, given by
\be
\Xi(\xi)&=&\sum_{x,y}w_0(x,y)e^{-\xi a_{x,y}},\\
p(x)&=&\frac 1{\Xi(\xi)}\sum_yw_0(x,y)e^{-\xi a_{x,y}},\\
w(x,y)&=&\frac {w_0(x,y)}{\Xi(\xi) p(x)}e^{-\xi a_{x,y}}.
\ee
Clearly, $a_{x,y}=a_{y,x}$ implies that the detailed balance condition is satisfied.
It is now straightforward to calculate the distributions $d_x(k)$. They are given by 
(\ref {gibbs}), with
\be
\Psi_\theta(k)
&=&\xi\sum_{x,y}k_{x,y}a_{x,y}+\sum_x\left(\sum_yk_{x,y}\right)\ln p(x)\cr
& &+n\ln\Xi(\xi).
\label {sec8temp}
\ee
A short calculation, using (\ref {sec8temp}) and the results of the section
on linear production of entropy, gives
\be
\ln\Xi(\xi)=S(p)+\frac 1nS^{(n)}_\theta-\xi \langle a\rangle.
\label {thermres}
\ee
Here, $S(p)$ is the Boltzmann-Gibbs entropy, and $\langle a\rangle$ is the path average of the matrix $a_{x,y}$.
Expression (\ref {thermres}) should be compared with the well-known result for the Massieu function
$\ln\Xi(\beta)=S-\beta U$, where $U=\langle H\rangle$ is the average energy.
This comparison shows that in (\ref {thermres}) the thermodynamic entropy is replaced by the sum of two contributions,
a static $S(p)$ and a dynamical $\frac 1nS^{(n)}_\theta$. 
Thermodynamic relations similar to (\ref {thermres}) are derived by Carati \cite {CA05,CA06}.
Here, we do not need a Legendre transformation to define entropy. In addition,
(\ref {sec8temp}) contains a path dependent term, which is absent in \cite {CA05,CA06}.

Related results are found in \cite {LAW05,LAW06}, where the parameter $s$,
appearing in the dynamical partition function,
is the argument of a Laplace transform. Its role is comparable to that of our parameter $\xi$,
which controls the transition probabilities of the Markov chain.

\paragraph* {A model of random walk}

Let us finally make the connection with the non-Markovian random walk model of \cite {VN05,VN06}.
The position after $n$ steps is denoted $X_n$. The increments $x_n=X_{n+1}-X_n$
are Markovian with transition matrix
\be
w=\left(\begin {array}{lr}\epsilon &1-\epsilon\\1-\mu &\mu\end {array}\right).
\ee
The two states are denoted $+$ and $-$. The enabling factors $w_0(x,y)$ are all equal to 1.

Let $X_0=0$. The position $X_n$ after $n$ steps is related to the record of transitions $k$
by $X_n=k_{++}+k_{-+}-k_{--}-k_{+-}$ (number of steps to the right minus number of steps
to the left). The energy of the polymer may be assumed to be proportional to
$k_{+-}+k_{-+}$, which is the number of changes of direction. Finally, let
$\Delta=k_{+-}-k_{-+}$ and note that $\Delta$ is either 0 or $\pm 1$.
Together with the identity $n=k_{++}+k_{-+}+k_{--}+k_{+-}$, this means that the record of
transitions $k$ can be expressed in terms of the physical quantities energy and position
of the end point, up to an error $\Delta$. This observation was used in \cite {VN05,VN06}
to prove that the joint probability
distribution of energy and position of endpoint automatically belongs to
the exponential family, be it with 3 parameters instead of 2.
As a consequence of the latter, the distribution is only approximately
that of Boltzmann-Gibbs, with a small error which is negligible in the
limit of large $n$.

\strut

\paragraph* {Discussion}

We have shown that some known results about Markov chains
can be formulated in terms of the distributions $d_x(k)$ which give the probability
of a transition record $k$ for paths of the Markov chain starting in state $x$.
These results are connected with what is known as the fluctuation theorem.
In addition we have pointed out, without going into much detail, that recent work of Carati \cite {CA05,CA06},
and of the present authors \cite {VN05,VN06} involves the same or related probability distributions.
We expect many more applications of our approach.
We believe that the distribution of transition records $d_x(k)$ will be the preferred tool,
rather than the fluctuation theorem (Theorem \ref {ft}).

Like in \cite {KJ98,CGE99}, the fluctuation theorem and the extensivity of the entropy production,
as proved here, hold for arbitrary number of steps $n$ because of the Markov assumption.
Usually, these results hold only in average. In particular, the assumption of a
Gibbsian field in space-time, as studied in \cite {MC99}, is more general than
Markovianity. A generalization of our approach to this context is unlikely
because it spoils the equivalence of paths that have the same transition record.

\strut
\begin {acknowledgments}

We thank Christian Maes and anonymous referees for helpful comments.
EVdS is Research Assistant of the Research Foundation - Flanders (FWO - Vlaanderen).

\end {acknowledgments}


\end{document}